\documentclass[conference]{IEEEtran}
\IEEEoverridecommandlockouts
\usepackage[linesnumbered,ruled,vlined]{algorithm2e}
\usepackage{booktabs}
\usepackage{multirow}
\usepackage{cite}
\usepackage{amsmath,amssymb,amsfonts}
\usepackage{graphicx}
\usepackage{textcomp}
\usepackage{xcolor}
\usepackage{geometry} 
\usepackage[linesnumbered,ruled,vlined]{algorithm2e}
\usepackage{cite}
\usepackage{amsmath}
\usepackage{array,color}
\usepackage{textcomp}
\usepackage{stfloats}
\usepackage{url}
\usepackage{verbatim}
\usepackage{graphicx}
\usepackage{cite}
\usepackage{orcidlink}
\usepackage{nomencl}
\usepackage{caption}
\usepackage{subcaption}
\usepackage[utf8]{inputenc}
\usepackage{pdfpages}
\usepackage{xcolor}
\usepackage{float}

\usepackage{float}

\begin{document}
\title{Advancing Healthcare: Innovative ML Approaches for Improved Medical Imaging in Data-Constrained Environments}

\author{
    \IEEEauthorblockN{
        Al Amin$^{1}$, Kamrul Hasan$^{1}$, Saleh Zein-Sabatto$^{1}$, Liang Hong$^{1}$, Sachin Shetty$^{2}$, Imtiaz Ahmed$^{3}$, \\Tariqul Islam$^{4}$
    }
    \IEEEauthorblockA{
        $^{1}$ Tennessee State University, Nashville, TN, USA \\
        $^{2}$ Old Dominion University, Norfolk, VA, USA \\
        $^{3}$ Howard University, Washington, DC, USA \\
        $^{4}$ Syracuse University, Syracuse, NY, USA \\
        Email: $\lbrace$\textit{aamin2, mhasan1, mzein, lhong}$\rbrace$@tnstate.edu, $\lbrace$\textit{sshetty}$\rbrace$@odu.edu, $\lbrace$\textit{imtiaz.ahmed}$\rbrace$@howard.edu, $\lbrace$\textit{mtislam}$\rbrace$@syr.edu
    }
}
\maketitle

\begin{abstract}
Healthcare industries face challenges when experiencing rare diseases due to limited samples. Artificial Intelligence (AI) communities overcome this situation to create synthetic data which is an ethical and privacy issue in the medical domain. This research introduces the CAT-U-Net framework as a new approach to overcome these limitations, which enhances feature extraction from medical images without the need for large datasets. The proposed framework adds an extra concatenation layer with downsampling parts, thereby improving its ability to learn from limited data while maintaining patient privacy. To validate, the proposed framework's robustness, different medical conditioning datasets were utilized including  COVID-19, brain tumors, and wrist fractures. The framework achieved nearly 98\% reconstruction accuracy, with a Dice coefficient close to 0.946. The proposed CAT-U-Net has the potential to make a big difference in medical image diagnostics in settings with limited data.
 
\end{abstract}


\section{Introduction}
\begin{figure*}[ht]
    \centering
    \includegraphics[height=6cm, width=\linewidth]{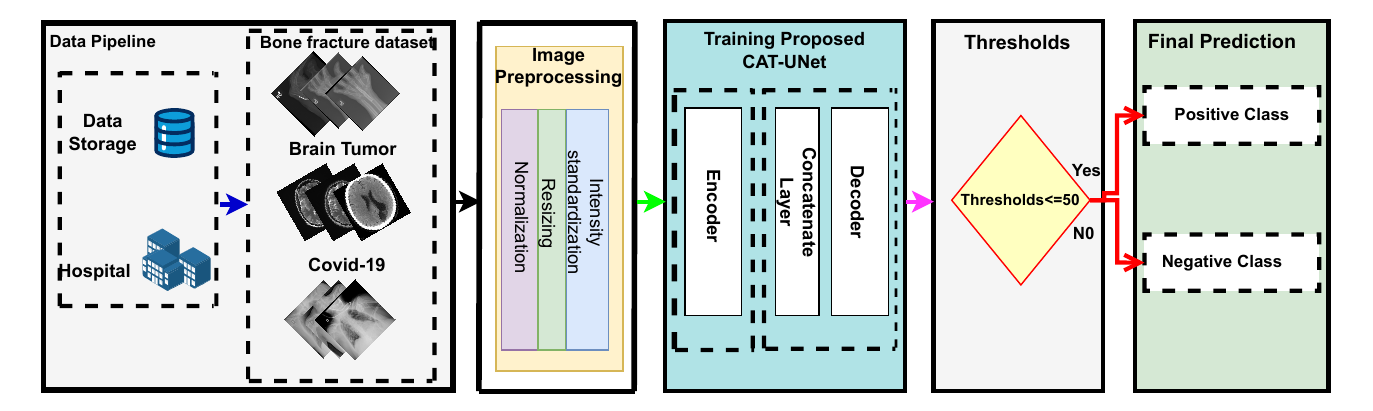}
     \caption{The pipeline of  proposed CAT-U-Net frameworks}
    \label{fig:cat-unet}
    \vspace{-4mm}
\end{figure*}

The ability to extract features from medical images strongly suits Deep Learning (DL) architectures \cite{9310359,li2021pathal,10437676, 10473196}. However, DL required large samples to reduce false positive and false negative rates \cite{boehringer2022active,liu2023structure, 10436915, 10454862,10582714}. Facing the challenge of data scarcity, especially in rare disease diagnostics, researchers and industry experts are increasingly turning to synthetic data generation and data augmentation techniques to bridge the gap \cite{chlap2021review,shorten2019survey,10464798}. In real-world clinical settings, AI models can be fooled by low-quality synthetic data when it comes to identifying the location and kind of diseases \cite{wang2023predictive, jaspers2024robustness}. Moreover, synthetic data generation in healthcare must navigate complex ethical and legal landscapes \cite{d2024synthetic,giuffre2023harnessing}, ensuring that privacy is preserved and inherent biases are not perpetuated \cite{10500144}. To overcome this limitation, the proposed framework reduces false positive and negative rates in limited dataset scenarios also improving the dice coefficient by over 94\%. Figure \ref{fig:cat-unet} represents the proposed CAT-Unet framework pipeline as an exemplary embodiment of innovation in medical imaging analysis, tailored to operate with high precision even when constrained to a minimal dataset of 100 positive samples. At the heart of its architecture lies a refined encoder-decoder framework: the encoder employs convolutional layers coupled with max pooling to distill critical features from images of brain tumors, COVID-19, and wrist fractures. After that, the decoder uses a new concatenation technique to improve and broaden these features; this merges the data's detailed and abstract properties, making the feature space much more robust. This method categorizes the decision according to a predetermined threshold value; scores of 50 or lower are regarded as positive, while scores exceeding 50 indicate negative outcomes.

In summary, the main contributions of this work are:
\begin{itemize}
    \item The work presents the CAT-U-Net, a new variant of the U-Net architecture with supplementary concatenation layers. This advancement enhances the network's ability to discern intricate features in medical imaging, significantly increasing the accuracy of detecting rare pathological manifestations.
    \item This research achieved a validation accuracy of 95\% and dice coefficient values of 0.946 by using several kinds of real-world medical imaging, such as MRI brain tumors, wrist fracture x-rays, and COVID-19 chest CT scans.
\end{itemize}

The remainder of this paper is structured as follows: Section II describes related works, and how different researchers solve data-constraints issues. Section III the proposed framework in detail. Section IV discusses dataset description, experimental setup, and result analysis  Section V highlights the summaries of this research and future research directions.

\section{Related Work}
Boosting the efficiency of DL methods in the face of limited data remains a crucial area of investigation. Conventional data augmentation practices are applicable in medical imaging \cite{10500428,10445413,10634882}. However, the generation of poor-quality synthetic data poses significant risks to medical research, and synthetic medical data can also raise privacy concerns \cite{10559993,10464345}. To address these challenges, Chen et al. proposed the Anatomy-Regularized Representation Learning-based Generative Adversarial Network (ARL-GAN) to address the challenges of limited labeled data and diverse imaging modalities in medical image segmentation. While effective in tasks like skull and cardiac substructure segmentation, ARL-GAN faces overfitting and difficulty generalizing across varied modalities \cite{chen2020anatomy}. Similarly, Shetty et al. introduced RAD-DCGAN to improve radiology image classification with synthetic data, achieving a 4-5\% accuracy boost over standard augmentation methods. However, RAD-DCGAN also struggles with ensuring data diversity and proper annotation, with risks to diagnostic reliability in real-world applications \cite{shetty2023data}.
Similarly, Mahmood et al. tackled the issue of data scarcity by proposing an innovative approach using unsupervised reverse domain adaptation and adversarial training. This method enables models trained on synthetic data to interpret real-world medical images accurately, significantly improving depth estimation accuracy in monocular endoscopy images. Despite its promise, this approach struggles to simulate the complex anatomical diversity of human tissues, leading to difficulties in generalizing models trained on synthetic data to real-world medical imaging, which may adversely impact accuracy \cite{mahmood2018unsupervised}. Rodriguez-Almeida et al. further explored synthetic data generation to address challenges associated with small and imbalanced medical datasets in disease prediction. Their framework showed that synthetic data can enhance classification performance, particularly in small datasets, but the results were mixed, with Gaussian Copula methods outperforming CTGANs in many cases. These findings highlight the limitations of relying solely on synthetic data, underscoring the need for approaches to improve diagnostic accuracy and maintain model reliability without over-reliance on synthetic data \cite{9851514}.
To tackle previous limitations, The suggested CAT-U-Net framework solves these problems by improving feature extraction and performing well across various datasets.

\section{Methodology}

\subsection{Proposed CAT-U-Net framework approach}
\begin{figure*}[h]
    \centering
    \includegraphics[width=\linewidth]{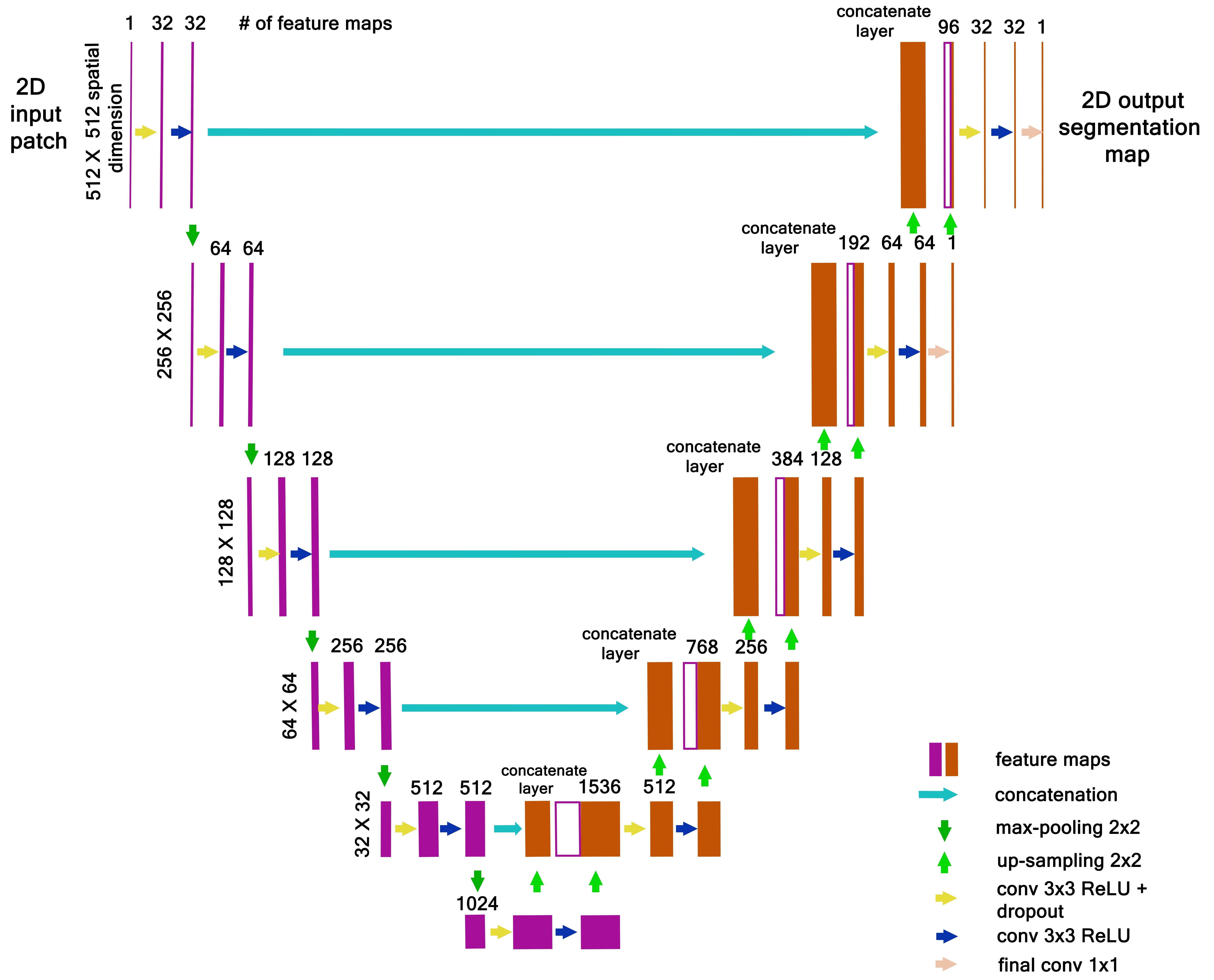}
     \caption{The proposed CAT-U-Net framework architecture with concatenation layers}
    \label{fig:my_labe2}
    \vspace{-4mm}
\end{figure*}

The proposed framework, depicted in Figure \ref{fig:my_labe2}, employs a symmetrical encoder-decoder structure enhanced with concatenation layers for improved feature extraction and diagnostic precision. The model architecture can be mathematically conceptualized in the following steps:

\subsubsection{Encoder}
The encoder is responsible for capturing the hierarchical features of the input medical images. For an input image \( I \in \mathbb{R}^{512 \times 512 \times C} \), the encoder applies a series of convolutional layers followed by max-pooling:

\begin{equation}
E_k(I) = P(\sigma(C(E_{k-1}(I))))
\end{equation}

where \( E_k(I) \) is the output of the \( k \)-th encoder block, \( C \) denotes a convolutional operation followed by a ReLU activation \( \sigma \), and \( P \) denotes a max-pooling operation.

\subsubsection{Concatenation Layers for Feature Extraction}
At each level of the decoder, features from the encoder are concatenated with upsampled features to preserve high-resolution details:

\begin{equation}
F_k = \text{Concat}(U(D_{k-1}), E_{n-k})
\end{equation}

Where \( F_k \) represents the concatenated features, \( U \) denotes the upsampling operation, \( D_{k-1} \) is the output of the previous decoder block, and \( E_{n-k} \) is the feature map from the encoder at the corresponding level.

\subsubsection{Decoder}
The decoder reconstructs the segmentation map from the compressed feature representation:

\begin{equation}
D_k(F) = \sigma(U(C(F_k)))
\end{equation}

Where \( D_k(F) \) is the output of the \( k \)-th decoder block, and the operations are defined as before.

\subsection{Robust Feature Learning with Concatenation Layers}
The introduction of concatenation layers in the CAT-U-Net framework fosters a more expressive feature representation, enhancing the learning capacity of the network even with limited data. The concatenation operation can be mathematically formalized as a union of feature sets from different levels of the network, enabling the preservation and utilization of multi-scale information:
\begin{equation}
\mathcal{F}_{concat}^{(l)} = \mathcal{F}_{encoder}^{(l)} \oplus \mathcal{U}(\mathcal{F}_{decoder}^{(l)})
\end{equation}
where \( \mathcal{F}_{concat}^{(l)} \) represents the concatenated feature map at layer \( l \), \( \mathcal{F}_{encoder}^{(l)} \) is the feature map from the encoder, \( \mathcal{F}_{decoder}^{(l)} \) is the feature map from the decoder, \( \mathcal{U} \) denotes the upsampling operation, and \( \oplus \) represents the concatenation operation.

\subsection{Segmentation Map Generation}
The final convolutional layer maps the concatenated features to the segmentation map:

\begin{equation}
S = \text{Conv}_{1\times1}(D_n(F))
\end{equation}

Where \( S \) represents the final segmentation map and \( \text{Conv}_{1\times1} \) denotes a 1x1 convolution operation that consolidates the feature maps into the final output.

\subsection{Binary Segmentation via Thresholding}
The model employs a thresholding technique to binarize the segmentation map, distinguishing between positive and negative diagnoses:

\begin{equation}
B(x,y) = 
  \begin{cases} 
    1 & \text{if } S(x,y) \leq T \\
    0 & \text{otherwise}
  \end{cases}
\end{equation}

Where \( B \) is the binarized segmentation map, \( (x,y) \) are the pixel coordinates, and \( T \) is the threshold value, empirically set to 50 in this study.

\subsection{Leveraging CAT-U-Net for Precise Medical Diagnosis: A Threshold-Based Approach}

This Algorithm 1 encapsulates the methodology employed by the CAT-U-Net model for diagnosing medical conditions from imaging data. The model exclusively trains on positive samples for a given condition, such as bone fractures. The initialization of the CAT-U-Net involves setting up an encoder-decoder architecture that includes concatenation layers to enhance feature extraction capabilities.
Once initialized, the model undergoes training with preprocessed positive samples, employing mean squared error (MSE) as the loss function to optimize its parameters. This training process leverages the model's ability to reconstruct the training images accurately, honing its capacity to identify intricate features indicative of the medical condition. The evaluation phase introduces a mixed test dataset comprising positive and negative samples. The CAT-U-Net predicts these samples, and the reconstruction performance is quantified using MSE losses. A thresholding method is then applied to these MSE losses: samples resulting in a loss equal to or less than 50 are classified as positive (indicating the presence of the condition), while those with losses greater than 50 are classified as negative.

\begin{algorithm}[ht]
\scriptsize
\caption{Proposed CAT-U-Net Based Medical Image Diagnosis Procedure}\label{alg:CATUNetDiagnosis}
\DontPrintSemicolon
\KwData{positive\_samples, mixed\_test\_samples}
\KwResult{final\_disease\_diagnosis}

\SetKwFunction{FMain}{Main}
\SetKwProg{Fn}{Function}{:}{\KwRet}

\Fn{\FMain{positive\_samples, mixed\_test\_samples}}{
    \textbf{Step 1: Initialize CAT-U-Net Model}\;
    cat\_u\_net $\gets$ initialize\_CAT\_U\_Net()\;
    
    \textbf{Step 2: Preprocess Positive Training Samples}\;
    preprocessed\_samples $\gets$ preprocess\_images(positive\_samples)\;
    
    \textbf{Step 3: Training the CAT-U-Net Model}\;
    X\_train, X\_test, y\_train, y\_test $\gets$ train\_test\_split(preprocessed\_samples)\;
    history $\gets$ cat\_u\_net.fit(X\_train, y\_train, validation\_data=(X\_test, y\_test))\;
    
    \textbf{Step 4: Model Evaluation with Mixed Test Samples}\;
    preprocessed\_test\_samples $\gets$ preprocess\_images(mixed\_test\_samples)\;
    predicted\_samples $\gets$ cat\_u\_net.predict(preprocessed\_test\_samples)\;
    
    \textbf{Step 5: Calculate Mean Squared Error (MSE) Loss}\;
    mse\_losses $\gets$ calculate\_mse(preprocessed\_test\_samples, predicted\_samples)\;
    
    \textbf{Step 6: Thresholding for Final Diagnosis}\;
    threshold $\gets$ 50\;
    final\_predictions $\gets$ []\;
    \ForEach{mse \textbf{in} mse\_losses}{
        \eIf{mse $\leq$ threshold}{
            final\_predictions.append('Positive')\;
        }{
            final\_predictions.append('Negative')\;
        }
    }
    
    \Return final\_predictions\;
    \textbf{print}("Final Disease Diagnosis:", final\_predictions)\;
}

\end{algorithm}
\DecMargin{1em}

\subsection{Optimization Framework for Limited Data}
The CAT-U-Net model is optimized to extract the maximum amount of diagnostic information from a limited dataset, employing a robust loss function combined with a regularization strategy to ensure a comprehensive learning of features. The optimization objective balances the need for accurate reconstruction against the complexity of the model, facilitating generalization from only 100 samples:

\begin{equation}
\theta^* = \arg\min_{\theta} \Big(L_{\text{MSE}}(\theta) + \lambda \mathcal{R}(\mathcal{F}_{concat}; \theta)\Big)
\end{equation}

Where \(L_{\text{MSE}}(\theta)\) is the mean squared error loss that measures the fidelity of image reconstruction, \( \mathcal{R} \) represents the regularization term that promotes the learning of a diverse feature set \( \mathcal{F}_{concat} \), and \( \lambda \) serves as a hyperparameter that ensures a balance between loss minimization and feature richness.

To achieve this objective, Stochastic Gradient Descent (SGD) is used to iteratively update the model parameters \( \theta \), reducing the loss on the training data:

\begin{equation}
\theta_{t+1} = \theta_t - \eta_t \cdot \nabla_{\theta} L_{\text{MSE}}(\theta_t)
\end{equation}

Here, \( \eta_t \) is the learning rate at epoch \( t \), dynamically adjusted according to a decay schedule contingent on the validation loss plateau, ensuring steady progress toward the optimal parameter set. This optimization framework is tailored for environments where data scarcity is a significant constraint, enabling the CAT-U-Net to perform with high diagnostic precision.

\subsection{Model Robustness and Healthcare Implications}
A theoretical framework supports the CAT-U-Net model's robustness, ensuring its performance with limited healthcare datasets governed by a specific boundedness condition on the feature space.
\begin{equation}
\|\mathcal{F}_{concat}\| \leq K
\end{equation}

Where \( K \) is a predefined constant that bounds the complexity of the concatenated feature space, ensuring the model captures a comprehensive set of features from medical images without succumbing to overfitting. This is crucial in healthcare settings where the variance in medical imagery is high, but the volume of data is often low. Coupled with a compact training set, this constraint ensures that CAT-U-Net maintains a high generalization capability, as demonstrated by its predictive accuracy on unseen data.

The optimization goal of the model encapsulates the balance between minimizing reconstruction error and maintaining feature diversity, thus effectively utilizing limited data:

\begin{align}
\theta^* &= \arg\min_{\theta} \Big(L_{\text{MSE}}(\theta) + \lambda \mathcal{R}(\mathcal{F}_{concat}; \theta)\Big), \\
&\quad \text{subject to } |D_{\text{train}}| \leq 100
\end{align}

Where \( \lambda \) is a hyperparameter that harmonizes the loss minimization and feature representation complexity. This equation articulates how CAT-U-Net is specifically designed to perform in environments where exhaustive data collection is challenging, a common scenario in medical diagnostics.

To finalize the diagnostic process, a threshold \( T \) on the MSE loss is used to discern positive from negative cases, encapsulating the diagnostic capability of CAT-U-Net within a decision boundary:

\begin{equation}
\text{Diagnosis}(x, y) = 
  \begin{cases} 
    \text{`Positive'} & \text{if } L_{\text{MSE}}(\hat{I}(x,y), I(x,y)) \leq \\T \\
    \text{`Negative'} & \text{otherwise}.
  \end{cases}
\end{equation}

This threshold-based diagnostic criterion aligns with the clinical need for accurate and reliable decision-making based on quantifiable image analysis.

The mathematical rigor presented herein demonstrates the robustness of the CAT-U-Net framework and its practical applicability in future healthcare industries, where data-driven decision-making is paramount. The ability to provide accurate diagnoses with fewer samples translates to quicker, more efficient medical care, potential reductions in healthcare costs, and broader accessibility of high-quality medical diagnostics.

\section{Proposed Framework Training and Optimization}
The CAT-U-Net framework was implemented using the TensorFlow framework due to its comprehensive support for DL operations and GPU acceleration. The following subsection explains this study's implementation, training procedure, and evaluation protocol.

\subsection{Training Procedure}
Training the CAT-U-Net involved steps, starting with preprocessing the input images. Each image was resized to a uniform dimension of \( 256 \times 256 \times 4 \) and normalized to have pixel values in the range [0, 1]. The network was then trained using the preprocessed images, with the Mean Squared Error (MSE) as the loss function. Stochastic Gradient Descent (SGD) with a learning rate of 0.01 was employed as the optimizer. To prevent overfitting, a dropout rate of 0.5 was applied after each convolutional layer in the decoder part of the network. The model was trained for 50 epochs with a batch size of 8, and the best-performing model on the validation set was saved for subsequent evaluation.

\subsection{Hyperparameter Optimization}
The model's hyperparameters were meticulously optimized to enhance performance, with a focus on the learning rate (\( \eta \)) adaptation strategy. To calibrate \( \eta \), an exponential decay function was applied, governed by the following relation:

\begin{equation}
\eta_{t+1} = \eta_{t} \times \gamma^{\lfloor\frac{t}{\tau}\rfloor}
\end{equation}

Where \( \eta_{t+1} \) is the learning rate for the subsequent epoch, \( \eta_{t} \) is the learning rate for the current epoch, \( \gamma \) is the decay rate, \( t \) is the current epoch number, and \( \tau \) is the patience parameter, denoting the number of epochs after which the plateau condition is evaluated. The decay rate \( \gamma \) was set to 0.1, and the learning rate was reduced if no improvement in validation loss was observed throughout ten epochs (\( \tau = 10 \)).

\section{RESULTS AND DISCUSSION}
\subsection{Experimental Setup}
Experiments were conducted on a High-Performance Computing (HPC) system featuring 31 GB of DDR4 RAM and an NVIDIA GeForce GTX 3070 GPU with 8 GB of GDDR6 memory, operating on a Linux Ubuntu environment. This setup was chosen to effectively manage the computational requirements of the CAT-U-Net framework, particularly for processing medical imaging data and training the model under data-constrained conditions. The framework's robustness was validated across MRI brain tumor images, COVID-19 chest CT scans, and wrist bone fracture X-ray images.

\subsection{Dataset Description}
The CAT-U-Net framework underwent validation using three distinct medical imaging datasets from Kaggle, each addressing different diagnostic challenges: the Br35H dataset for MRI brain tumor detection \cite{18}, a chest CT scan dataset for COVID-19 identification \cite{20}, and an X-ray dataset for Wrist Bone Fracture  \cite{19}. Training utilized 100 images from each dataset, while testing employed 50 images, demonstrating the framework’s robustness in detecting various pathological features, particularly in scenarios with limited data, a common challenge in medical diagnostics.

\subsection{Performance Metrics for Evaluation}
The proposed CAT-U-Net framework's effectiveness was assessed using reconstruction accuracy and the Dice coefficient to measure image precision and segmentation under limited data constraints.

\subsubsection{Reconstruction Accuracy}
Reconstruction accuracy was computed as the inverse of the mean squared error (MSE) between the original and reconstructed images across the test set. This metric reflects the model's ability to faithfully reconstruct key features in medical images, which is crucial for accurate diagnosis:
\begin{equation}
\text{Reconstruction Accuracy} = 1 - \frac{1}{N} \sum_{i=1}^{N} \|I_i - \hat{I}_i\|^2
\end{equation}
where \(I_i\) represents the original image, \(\hat{I}_i\) is the reconstructed image by the model, and \(N\) is the number of images in the test dataset.
\subsubsection{Dice Coefficient}
The Dice coefficient, also known as the Dice similarity coefficient (DSC), was used to measure the overlap between the predicted segmentation and the ground truth, providing a spatial accuracy assessment of the segmentation:

\begin{equation}
\text{Dice Coefficient} = \frac{2 \times |X \cap Y|}{|X| + |Y|}
\end{equation}

Where \(X\) represents the set of pixels in the ground truth segmentation, and \(Y\) denotes the set of pixels in the predicted segmentation by the model. A higher Dice score indicates more significant overlap and, thus, higher segmentation accuracy.

\subsection{Quantitative Evaluation of CAT-U-Net Across Diverse Medical Imaging Datasets}

\begin{figure}[h]
    \centering
    \includegraphics[width=\linewidth]{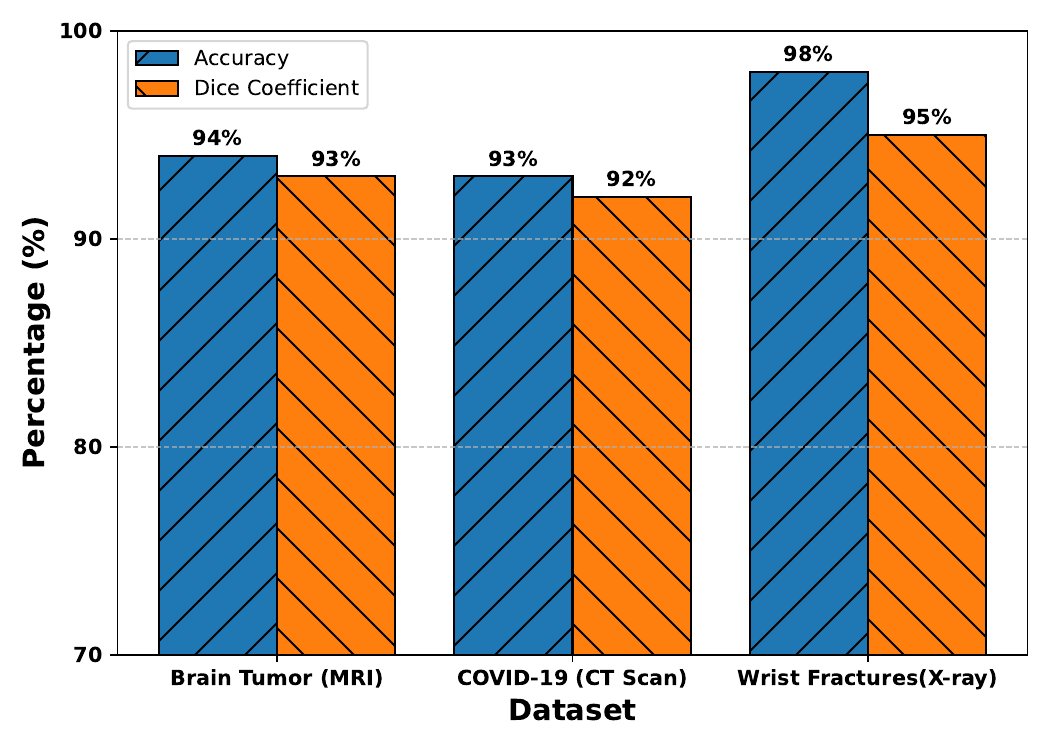}
     \caption{Performance Analysis of the Proposed Framework Across Different Datasets}
    \label{fig:my_result}
    \vspace{-4mm}
\end{figure}

\begin{figure}[h]
    \centering
    \includegraphics[width=\linewidth]{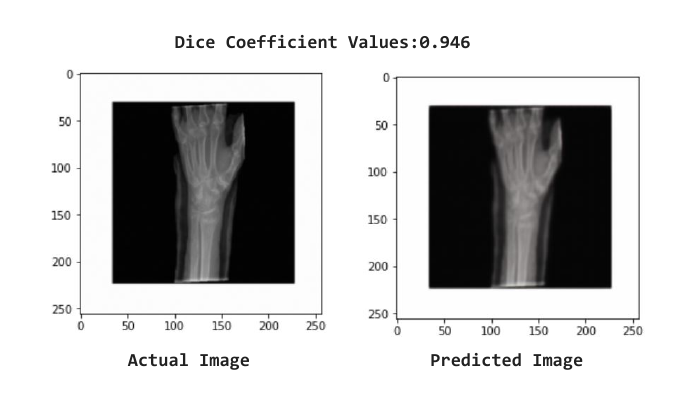}
     \caption{Dice-Cofficient value for Bone fracture dataset}
    \label{fig:my_label10}
    \vspace{-4mm}
\end{figure}

\begin{figure}[h]
    \centering
-    \includegraphics[height=7cm, width=\linewidth]{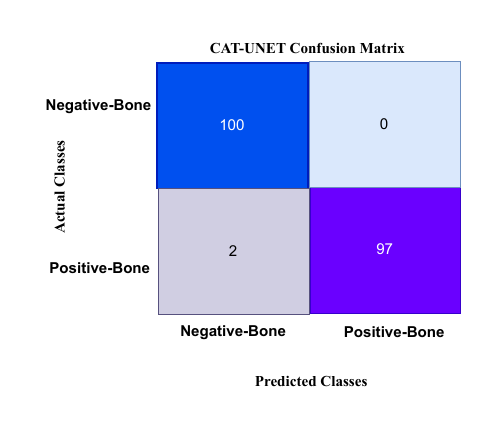}
     \caption{Proposed framework Confusion matrix}
    \label{fig:my_label11}
    \vspace{-4mm}
\end{figure}

\begin{figure}[h]
    \centering
    \includegraphics[height=5cm, width=\linewidth]{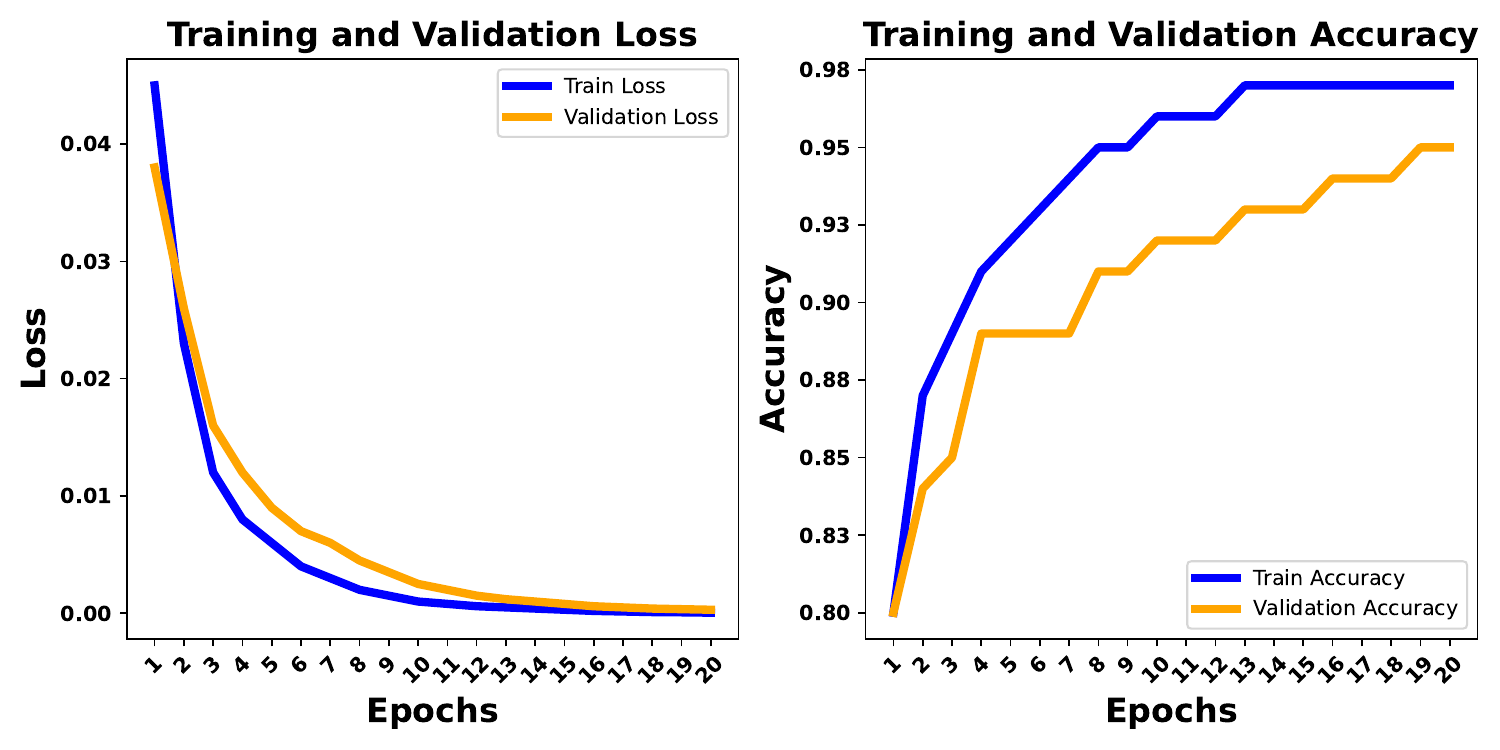}
     \caption{Training and Validation loss of propose framework}
    \label{fig:my_label12}
    \vspace{-4mm}
\end{figure}

The CAT-U-Net framework demonstrates high efficacy in medical image analysis, as evidenced by comprehensive statistical and visual evaluations across multiple datasets. In the Bone Fracture dataset, a confusion matrix (Figure \ref{fig:my_label11}) reveals the model's high predictive accuracy, correctly diagnosing 97 out of 99 positive cases, showcasing its sensitivity and specificity in detecting medical conditions. This accuracy is further supported by a high Dice coefficient of 0.946 (Figure \ref{fig:my_label10}), indicating an exceptional overlap between the model's predicted segmentations and the ground truth. Additionally, the training and validation loss and accuracy curves (Figure \ref{fig:my_label12}) exhibit a steady convergence, highlighting the model's robust learning capability from a limited number of training samples. As depicted in Figure \ref{fig:my_result}, the CAT-U-Net achieves consistently high performance across different datasets, with reconstruction accuracies of 94\% for Brain Tumor (MRI), 93\% for COVID-19 (CT Scan), and 98\% for Wrist Fractures (X-ray). The corresponding Dice coefficients also remain high, at 93\%, 92\%, and 95\%, respectively. These results confirm the model's adaptability and precision across diverse clinical scenarios, underscoring its potential as a reliable tool for medical image diagnostics.

\subsection{Quantitative Comparison with Existing State-of-the-Art Methods}

\begin{figure}[h]
    \centering
    \includegraphics[height=7.5cm, width=\linewidth]{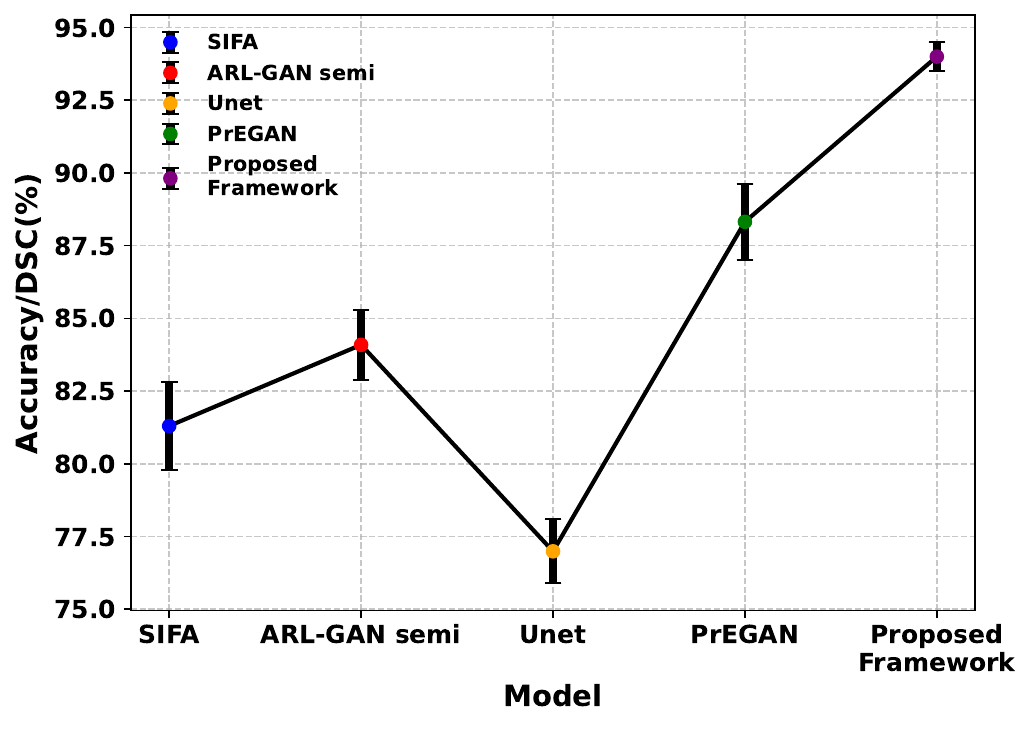}
     \caption{Comparative Performance Analysis of Proposed and Existing Models}
    \label{fig:my_comp}
    \vspace{-4mm}
\end{figure}

Figure \ref{fig:my_comp} provides a comparative analysis of the proposed CAT-U-Net framework against several state-of-the-art models, including SIFA \cite{chen2020anatomy}, ARL-GAN semi-supervised \cite{chen2020anatomy}, U-Net \cite{ronneberger2015u}, and PrEGAN \cite{10500428}, in terms of accuracy o Dice Similarity Coefficient (DSC). The proposed framework achieves superior performance with a DSC of 94.0\%, outperforming other methods such as PrEGAN (88.32\%), ARL-GAN semi-supervised (84.09\%), SIFA (81.3\%), and U-Net (76.99\%). These results demonstrate the efficacy of the CAT-U-Net framework in achieving higher segmentation accuracy and robustness, particularly in medical imaging tasks with limited data availability. 

\section{CONCLUSION AND FUTURE WORK}
In conclusion, while the CAT-U-Net framework presents a significant advancement in medical image analysis, its broader adoption in the healthcare industry will depend on addressing integration challenges and ensuring robustness across diverse clinical environments. Future research could focus on optimizing the framework for even smaller datasets and more complex imaging modalities and integrating advanced adversarial training methods to enhance robustness against adversarial attacks in clinical applications. Continued research in these areas will be essential to unlocking the full potential of this technology for improving patient outcomes.

\section{ACKNOWLEDGMENT}
This material is based on work supported by the National
Science Foundation Award Numbers 2205773 and 2219658.
\bibliographystyle{ieeetr}
\bibliography{references.bib}

\begin{thebibliography}{10}

\bibitem{9310359}
V.~Nath, D.~Yang, B.~A. Landman, D.~Xu, and H.~R. Roth, ``Diminishing uncertainty within the training pool: Active learning for medical image segmentation,'' {\em IEEE Transactions on Medical Imaging}, vol.~40, no.~10, pp.~2534--2547, 2021.

\bibitem{li2021pathal}
W.~Li, J.~Li, Z.~Wang, J.~Polson, A.~E. Sisk, D.~P. Sajed, W.~Speier, and C.~W. Arnold, ``Pathal: An active learning framework for histopathology image analysis,'' {\em IEEE transactions on medical imaging}, vol.~41, no.~5, pp.~1176--1187, 2021.

\bibitem{10437676}
M.~S. Ahmed and S.~Giordano, ``Pre-trained lightweight deep learning models for surgical instrument detection: Performance evaluation for edge inference,'' in {\em GLOBECOM 2023 - 2023 IEEE Global Communications Conference}, pp.~3873--3878, 2023.

\bibitem{10473196}
S.~Pattanaik, C.~Chakraborty, S.~Behera, S.~K. Majhi, and S.~K. Pani, ``An miot framework of consumer technology for medical diseases prediction,'' {\em IEEE Transactions on Consumer Electronics}, vol.~70, no.~1, pp.~3754--3761, 2024.

\bibitem{boehringer2022active}
A.~S. Boehringer, A.~Sanaat, H.~Arabi, and H.~Zaidi, ``An active learning approach to deep learning glioma segmentation from brain mr images,'' in {\em 2022 IEEE Nuclear Science Symposium and Medical Imaging Conference (NSS/MIC)}, pp.~1--2, IEEE, 2022.

\bibitem{liu2023structure}
S.~Liu, S.~Yin, L.~Qu, M.~Wang, and Z.~Song, ``A structure-aware framework of unsupervised cross-modality domain adaptation via frequency and spatial knowledge distillation,'' {\em IEEE Transactions on Medical Imaging}, 2023.

\bibitem{10436915}
A.~Farrag, G.~Gad, Z.~M. Fadlullah, and M.~M. Fouda, ``Mammogram tumor segmentation with preserved local resolution: An explainable ai system,'' in {\em GLOBECOM 2023 - 2023 IEEE Global Communications Conference}, pp.~314--319, 2023.

\bibitem{10454862}
E.~Haque, K.~Hasan, I.~Ahmed, M.~S. Alam, and T.~Islam, ``Towards an interpretable ai framework for advanced classification of unmanned aerial vehicles (uavs),'' in {\em 2024 IEEE 21st Consumer Communications \& Networking Conference (CCNC)}, pp.~644--645, 2024.

\bibitem{10582714}
R.~Wiencek and S.~Ghosh, ``Deep reinforcement learning for adaptive optimization of pi control for microgrid under fault and variable loading,'' in {\em 2024 6th Global Power, Energy and Communication Conference (GPECOM)}, pp.~826--831, 2024.

\bibitem{chlap2021review}
P.~Chlap, H.~Min, N.~Vandenberg, J.~Dowling, L.~Holloway, and A.~Haworth, ``A review of medical image data augmentation techniques for deep learning applications,'' {\em Journal of Medical Imaging and Radiation Oncology}, vol.~65, no.~5, pp.~545--563, 2021.

\bibitem{shorten2019survey}
C.~Shorten and T.~M. Khoshgoftaar, ``A survey on image data augmentation for deep learning,'' {\em Journal of big data}, vol.~6, no.~1, pp.~1--48, 2019.

\bibitem{10464798}
A.~Amin, K.~Hasan, S.~Zein-Sabatto, D.~Chimba, I.~Ahmed, and T.~Islam, ``An explainable ai framework for artificial intelligence of medical things,'' in {\em 2023 IEEE Globecom Workshops (GC Wkshps)}, pp.~2097--2102, 2023.

\bibitem{wang2023predictive}
Z.~Wang, {\em Predictive Learning from Real-World Medical Data: Overcoming Quality Challenges}.
\newblock PhD thesis, 2023.

\bibitem{jaspers2024robustness}
T.~J. Jaspers, T.~G. Boers, C.~H. Kusters, M.~R. Jong, J.~B. Jukema, A.~J. de~Groof, J.~J. Bergman, P.~H. de~With, and F.~van~der Sommen, ``Robustness evaluation of deep neural networks for endoscopic image analysis: Insights and strategies,'' {\em Medical Image Analysis}, p.~103157, 2024.

\bibitem{d2024synthetic}
G.~D'Acquisto, ``Synthetic data and data protection laws,'' {\em Journal of Data Protection \& Privacy}, vol.~6, no.~3, pp.~227--239, 2024.

\bibitem{giuffre2023harnessing}
M.~Giuffr{\`e} and D.~L. Shung, ``Harnessing the power of synthetic data in healthcare: innovation, application, and privacy,'' {\em NPJ Digital Medicine}, vol.~6, no.~1, p.~186, 2023.

\bibitem{10500144}
A.~Amin, K.~Hasan, S.~Zein-Sabatto, D.~Chimba, L.~Hong, I.~Ahmed, and T.~Islam, ``Empowering healthcare through privacy-preserving mri analysis,'' in {\em SoutheastCon 2024}, pp.~1534--1539, 2024.

\bibitem{10500428}
S.~T. Ahmed, R.~Sivakami, V.~K. V, M.~T. R, S.~B. Khan, A.~Mashat, and A.~Almusharraf, ``Pregan: Privacy enhanced clinical emr generation: Leveraging gan model for customer de-identification,'' {\em IEEE Transactions on Consumer Electronics}, pp.~1--1, 2024.

\bibitem{10445413}
S.~Islam, M.~T. Aziz, H.~R. Nabil, J.~R. Jim, M.~F. Mridha, M.~M. Kabir, N.~Asai, and J.~Shin, ``Generative adversarial networks (gans) in medical imaging: Advancements, applications, and challenges,'' {\em IEEE Access}, vol.~12, pp.~35728--35753, 2024.

\bibitem{10634882}
A.~Amin, K.~Hasan, and M.~S. Hossain, ``Xai-empowered mri analysis for consumer electronic health,'' {\em IEEE Transactions on Consumer Electronics}, pp.~1--1, 2024.

\bibitem{10559993}
S.~Abdel-Khalek, A.~D. Algarni, G.~Amoudi, S.~Alkhalaf, F.~M. Alhomayani, and S.~Kathiresan, ``Leveraging ai-generated content for synthetic electronic health record generation with deep learning-based diagnosis model,'' {\em IEEE Transactions on Consumer Electronics}, pp.~1--1, 2024.

\bibitem{10464345}
S.~A. Naeini, L.~Simmatis, D.~Jafari, Y.~Yunusova, and B.~Taati, ``Improving dysarthric speech segmentation with emulated and synthetic augmentation,'' {\em IEEE Journal of Translational Engineering in Health and Medicine}, vol.~12, pp.~382--389, 2024.

\bibitem{chen2020anatomy}
X.~Chen, C.~Lian, L.~Wang, H.~Deng, T.~Kuang, S.~Fung, J.~Gateno, P.-T. Yap, J.~J. Xia, and D.~Shen, ``Anatomy-regularized representation learning for cross-modality medical image segmentation,'' {\em IEEE transactions on medical imaging}, vol.~40, no.~1, pp.~274--285, 2020.

\bibitem{shetty2023data}
S.~Shetty, V.~Ananthanarayana, and A.~Mahale, ``Data augmentation vs. synthetic data generation: An empirical evaluation for enhancing radiology image classification,'' in {\em 2023 IEEE 17th International Conference on Industrial and Information Systems (ICIIS)}, pp.~1--6, IEEE, 2023.

\bibitem{mahmood2018unsupervised}
F.~Mahmood, R.~Chen, and N.~J. Durr, ``Unsupervised reverse domain adaptation for synthetic medical images via adversarial training,'' {\em IEEE transactions on medical imaging}, vol.~37, no.~12, pp.~2572--2581, 2018.

\bibitem{9851514}
A.~J. Rodriguez-Almeida, H.~Fabelo, S.~Ortega, A.~Deniz, F.~J. Balea-Fernandez, E.~Quevedo, C.~Soguero-Ruiz, A.~M. Wägner, and G.~M. Callico, ``Synthetic patient data generation and evaluation in disease prediction using small and imbalanced datasets,'' {\em IEEE Journal of Biomedical and Health Informatics}, vol.~27, no.~6, pp.~2670--2680, 2023.

\bibitem{18}
A.~Hamada, ``Brain tumor detection,'' 2021.
\newblock Accessed: 2023-04-08.

\bibitem{20}
T.~RAHMAN, D.~M. Chowdhury, and A.~Khandakar, ``Covid-19 radiography database,'' 2022.
\newblock Accessed: 2024-04-12.

\bibitem{19}
T.~V.~A. RAJ, ``Wrist bone fracture segmentation,'' 2021.
\newblock Accessed: 2024-04-12.

\bibitem{ronneberger2015u}
O.~Ronneberger, P.~Fischer, and T.~Brox, ``U-net: Convolutional networks for biomedical image segmentation,'' in {\em Medical image computing and computer-assisted intervention--MICCAI 2015: 18th international conference, Munich, Germany, October 5-9, 2015, proceedings, part III 18}, pp.~234--241, Springer, 2015.

\end{thebibliography}
\end{document}